\documentclass[aps,prl,twocolumn,groupedaddress,showpacs]{revtex4-1}

\usepackage{graphicx}
\usepackage{dcolumn} 
\usepackage{bm}      
\usepackage{soul}    
\usepackage{pdfpages}

\begin{document}

\title{Measurement of the Homogeneous Contact of a Unitary Fermi gas}
\author{Yoav Sagi, Tara E. Drake, Rabin Paudel, and Deborah S. Jin}
\email[Electronic address: ]{jin@jilau1.colorado.edu}

\affiliation{JILA, National Institute of Standards and Technology and the University of Colorado, and the Department of Physics, University of Colorado, Boulder, CO 80309-0440, USA}

\date{\today} \begin{abstract}

By selectively probing the center of a trapped gas, we measure the local, or homogeneous, contact of a unitary Fermi gas as a function of temperature.  Tan's contact, $C$, is proportional to the derivative of the energy with respect to the interaction strength, and is thus an essential thermodynamic quantity for a gas with short-range correlations. Theoretical predictions for the temperature dependence of $C$ differ substantially, especially near the superfluid transition, $T_c$, where $C$ is predicted to either sharply decrease, sharply increase, or change very little. For $T/T_F>0.4$, our measurements of the homogeneous gas contact show a gradual decrease of $C$ with increasing temperature, as predicted by theory. We observe a sharp decrease in $C$ at $T/T_F=0.16$, which may be due to the superfluid phase transition.  While a sharp decrease in C below $T_c$ is predicted by some many-body theories, we find that none of the predictions fully accounts for the data.

\end{abstract}


\maketitle

The collective behavior of an ensemble of strongly interacting fermions is central to many physical systems including liquid $^3$He, high-T$_c$ superconductors, quark-gluon plasma, neutron stars, and ultracold Fermi gases. However, theoretical understanding of strongly interacting fermions is challenging due to the many-body nature of the problem and the fact that there is no obvious small parameter for a perturbative analysis. Therefore, in order to establish the validity and applicability of theoretical approaches, it is essential to compare them against experimental results. Ultracold atomic Fermi gases are ideal for this purpose, as they provide excellent controllability, reproducibility, and unique detection methods \cite{Regal2006,Ketterle2008}. In particular, changing the magnetic field in the vicinity of a Feshbach resonance enables precise control of the interactions, which are characterized by the s-wave scattering length \cite{RevModPhys.82.1225}. On resonance, the scattering length diverges and the behavior of the unitary gas no longer depends on it. Testing theories in this regime is especially desirable.

An outstanding issue for the unitary Fermi gas is the nature of the normal state just above the transition temperature, T$_c$, for a superfluid of paired fermions. Some theories of strongly interacting Fermi gases (BCS-BEC crossover theories) predict that the normal state is not the ubiquitous Fermi liquid but instead involves incoherent fermion pairing (preformed pairs) in what has been termed the pseudogap state \cite{Chen20051}. It has been suggested that the pseudogap state affects the temperature dependence of a quantity called Tan's contact \cite{PhysRevA.82.021605-2010}. The contact, which is a measure of the short-range correlation function, has been shown to be an essential thermodynamic parameter for ensembles with short-range interactions \cite{Tan08b,Tan08,Tan08a,Braaten08,Zhang2009,Braaten2012}. The contact connects many seemingly unrelated quantities through a set of universal relations that are valid for any temperature, any interaction strength, and any phase of the system.  While the value of the contact, as well as many of these relations, were tested successfully at low temperature \cite{PhysRevLett.104.235301,PhysRevLett.105.070402,PhysRevLett.95.020404,Castin2009}, there are significant discrepancies among theories on how the contact of a unitary homogeneous Fermi gas depends on temperature, especially around T$_c$ \cite{PhysRevA.80.023615-2009,PhysRevA.82.021605-2010,Enss2011770-2011,1367-2630-13-3-035007-2011,PhysRevLett.106.205302}. The temperature dependence of the contact was recently measured for a trapped unitary Fermi gas \cite{PhysRevLett.106.170402}. However, for the trapped gas, averaging over the inhomogeneous density distribution washes out any temperature-dependent features, and the measurement was unable to differentiate between theoretical models. Here we present a measurement of the homogeneous contact, which can be directly compared to the predictions of different many-body theories.

\begin{figure}
\center\includegraphics[width=8cm]{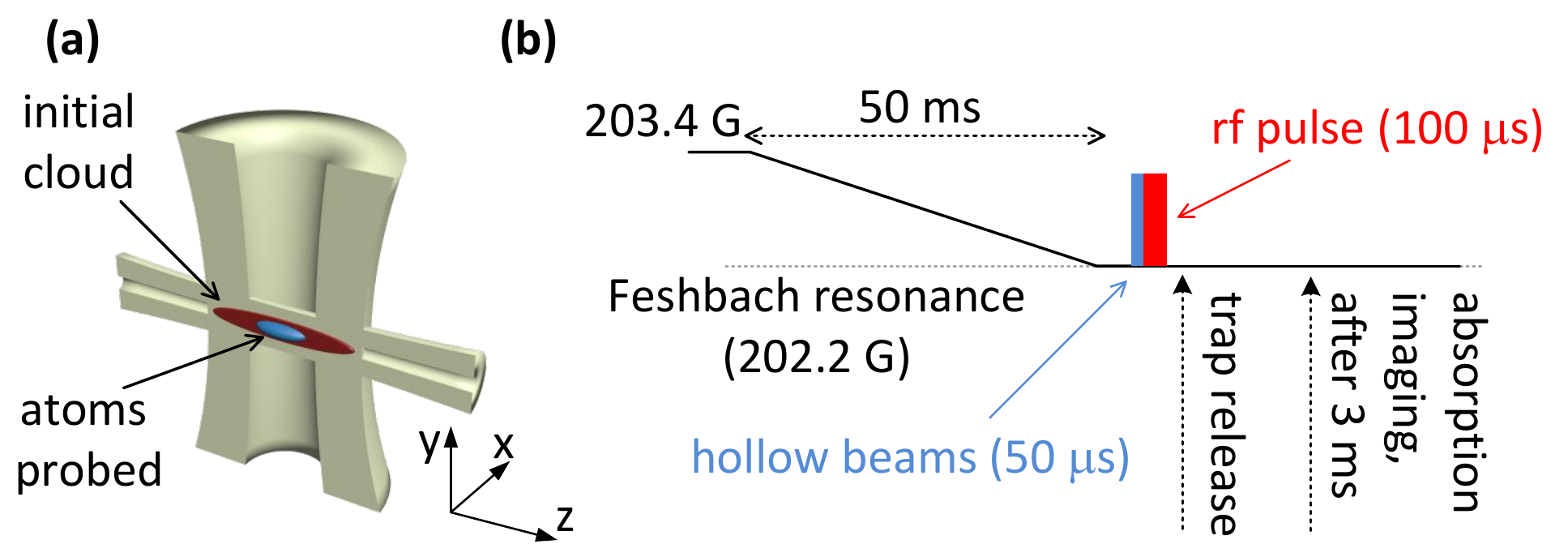}
    \caption{Schematics of the experiment. \textbf{(a)} We probe the center of the gas by optically pumping the outer parts of the cigar-shaped cloud to a dark state using two intersecting second-order Laguerre-Gaussian beams \cite{PhysRevA.86.031601}. By changing the beams' power, we control the fraction of atoms probed. The power of the two beams is set such that the number of atoms optically pumped by each beam is about the same. \textbf{(b)} The magnetic field is ramped from $203.4$ G, where the atoms are initially prepared, to the Feshbach resonance. The hollow light beams are turned on $280\, \mu$s before trap release; initially, the beam that propagates perpendicular to the long axis of the cloud is pulsed on for $10\, \mu $s followed by $40\, \mu$s of the second beam. The line shape is measured using an rf pulse with a total duration of $100\, \mu$s and a gaussian field envelope with $\sigma=17 \, \mu s$, centered $180\, \mu$s before trap release. The cloud expands for $3$ ms before being detected by absorption imaging. To improve the signal-to-noise ratio, we remove the remaining atoms from the $|9/2,-9/2\rangle$ and $|9/2,-7/2\rangle$ states and then transfer the outcoupled atoms in the $|9/2,-5/2\rangle$ state to the $|9/2,-9/2\rangle$ state, where we image on the cycling transition \cite{Gaebler2010}.}\label{schematics}
\end{figure}

We perform the experiments with an optically trapped ultracold gas of $^{40}\rm{K}$ atoms in an equal mixture of the $|f,m_f\rangle= |9/2, -9/2\rangle$ and
$|9/2,-7/2\rangle$ spin states, where $f$ is the quantum number denoting the total atomic spin and $m_f$ is its projection \cite{Stewart2008}. We determine the contact by combining rf spectroscopy with a recently demonstrated technique to probe the local properties of a trapped gas \cite{PhysRevA.86.031601}. Probing atoms locally is accomplished by intersecting two perpendicularly propagating hollow light beams that optically pump atoms at the edge of the cloud into a spin state that is dark to the detection (see figure \ref{schematics}a). The experimental sequence is depicted in figure \ref{schematics}b. The magnetic field is ramped adiabatically to the Feshbach resonance and kept at this value for $2$ ms before abruptly shutting off the trapping potential. Before the potential is shut off, the hollow beams are pulsed on, followed by the rf pulse, which transfers a small fraction of the atoms in the occupied $|9/2,-7/2\rangle$ state to the initially unoccupied $|9/2,-5/2\rangle$ state (which is weakly interacting with the other two spin states). We detect these atoms using absorption imaging after 3 ms of expansion. The temperature of the gas is varied by changing the final depth of the optical dipole trap in the evaporation process \cite{Stewart2008}. The number of atoms per spin state after the evaporation ranges from $50,000$ to $220,000$. For the data presented in this paper, the radial trapping frequency, $\omega_r$, ranges from $2\pi\times 200$ Hz to $2\pi\times 410$ Hz, while the axial trapping frequency, $\omega_z$, ranges from $2\pi\times 19$ Hz to $2\pi\times 25$ Hz.

\begin{figure}
\center\includegraphics[width=8cm]{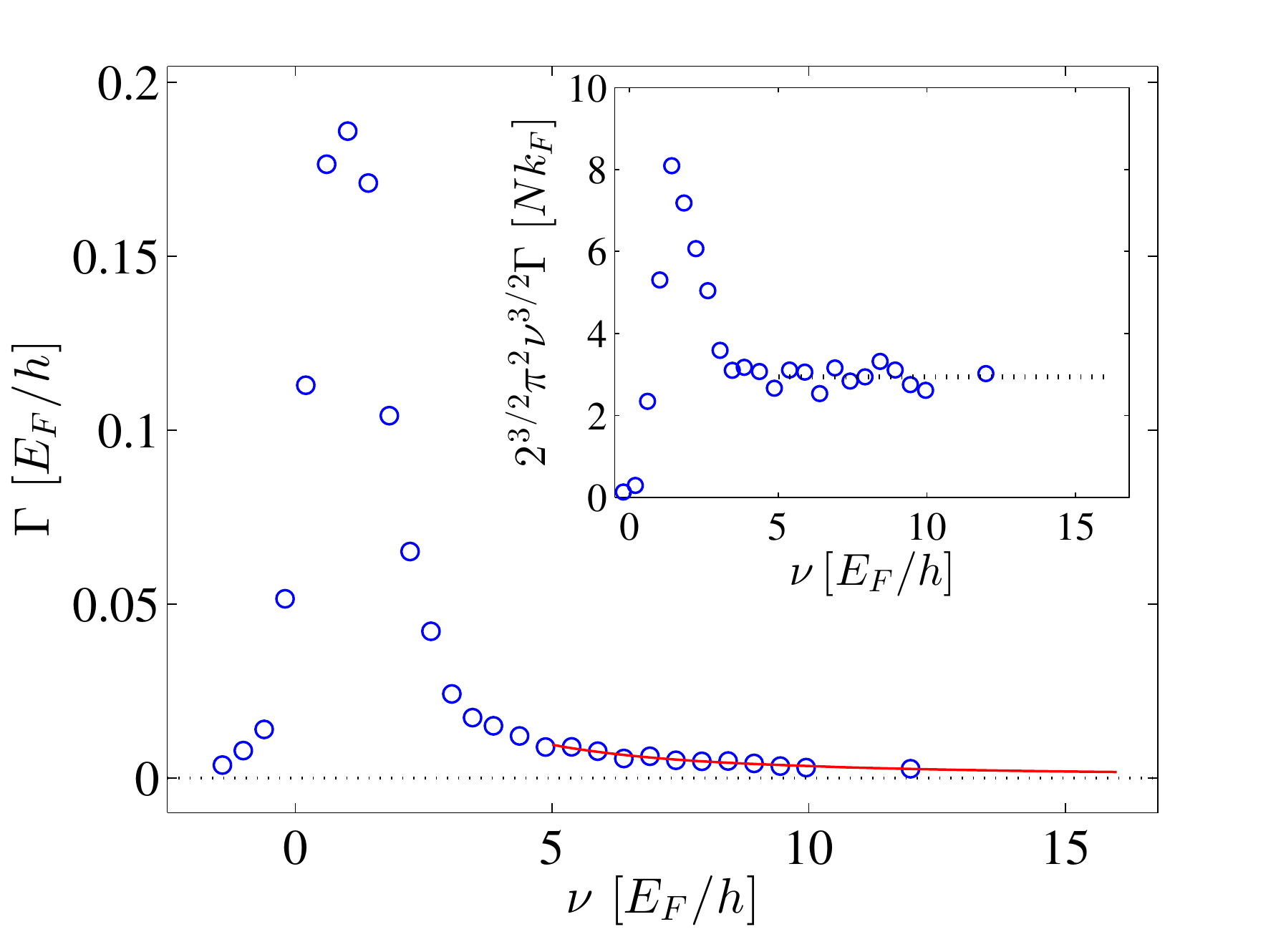}
    \caption{An rf line shape for the unitary Fermi gas at $T/T_F=0.25$ with $30\%$ of the atoms probed. The solid (red) line is a fit to Eq.(\ref{Tan_relation}) with the normalization $\int_{-\infty}^\infty \Gamma(\nu) \mathrm{d}\nu=1/2$, due to the $50\%-50\%$ spin mixture. The inset shows the same data multiplied by $2^{3/2} \pi^2 \nu^{3/2}$. We make sure the rf pulse induces only a small perturbation, by setting its power to well below the value where we see the onset of saturation of the number of outcoupled atoms \cite{EPAPS_note2}. The measurement at different frequencies is done with different rf powers, and when analyzing the data, we linearly scale the measured number of atoms outcoupled at each frequency to correspond to a common rf power.}\label{raw_data}
\end{figure}

The contact is extracted from a measurement of the rf line shape $\Gamma(\nu)$ \cite{PhysRevLett.104.235301}, where $\Gamma(\nu)$ is the rate of atoms transferred from one of the two interacting spin states to a third state, by an rf pulse centered at a frequency detuning $\nu$. A representative data set, where the hollow light beams were used to select the central $30\%$ of the atom cloud, is shown in figure \ref{raw_data}. For each line shape, we take data at 30 different detunings between $-16$ kHz and $+116$ kHz, where $\nu=0$ is defined as the single-particle transition frequency between the $|9/2,-7/2\rangle$ and $|9/2,-5/2\rangle$ states (measured for a spin polarized gas in the $|9/2,-7/2\rangle$ state). The highest frequency typically corresponds to approximately $13\, E_F/h$. The high frequency tail of the rf line shape is predicted to scale as $\nu^{-3/2}$, with the contact connecting the amplitude of the high frequency tail through (see Ref. \cite{Braaten2012} and references therein):
\begin{equation}\label{Tan_relation}
\frac{\Gamma(\nu)}{\int_{-\infty}^\infty \Gamma(\nu') \mathrm{d}\nu'}=\frac{C/(N k_F)}{\sqrt{2}\pi^2\nu^{3/2}} \ \ \ \rm{for} \ \nu\rightarrow \infty
\end{equation}
where $N$ is the total number of atoms, and $\hbar k_F$ is the Fermi momentum, and $\nu$ is the rf detuning in units of the Fermi energy, $E_F/h$, with $h$ being the Planck constant ($2\pi\hbar\equiv h$). The inset of figure \ref{raw_data} shows $\Gamma(\nu)$ multiplied by $2^{3/2} \pi^2 \nu^{3/2}$, where we observe a plateau for frequencies higher than $5\, E_F/h$. We extract the contact by fitting the measured $\Gamma(\nu)$ for $\nu>5\, E_F/h$ to Eq.(\ref{Tan_relation}) (solid line in figure \ref{raw_data}). For the normalization, we integrate the line shape, including the tail, up to $\nu=\hbar/m r_{\rm{eff}}^2$, where $r_{\rm{eff}}$ is the effective range of the interaction \cite{RevModPhys.82.1225} (which is approximately $300 E_F/\hbar$).

\begin{figure}
\center\includegraphics[width=8cm]{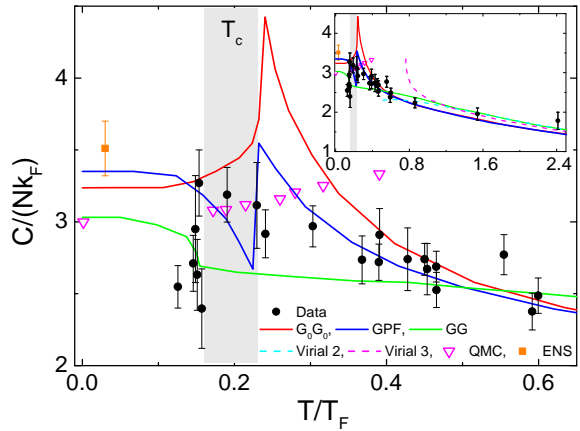}
    \caption{The contact of a nearly homogeneous sample (about $30\%$ of the trapped atoms probed), versus $T/T_F$ at unitarity (black circles). The shaded area marks the superfluid phase transition, with some uncertainty in its exact position ($T_c/T_F=0.16-0.23$) \cite{PhysRevA.75.023610}. As a comparison, we plot the gaussian pair-fluctuation NSR model (GPF) \cite{1367-2630-13-3-035007-2011}, the self-consistent t-matrix model (GG) \cite{PhysRevA.75.023610}, the non-self-consistent t-matrix model (G$_0$G$_0$) \cite{PhysRevA.82.021605-2010}, the 2nd and 3rd order virial expansion \cite{1367-2630-13-3-035007-2011}, a quantum Monte-Carlo calculation (QMC) \cite{PhysRevLett.106.205302}, and the contact extracted from a thermodynamic measurement done at ENS \cite{Navon07052010}. The error bars represent one standard deviation. The inset shows the high temperature behavior of the contact, where we find good agreement with the virial expansion.}\label{contact_vs_temp}
\end{figure}

The main result of the paper, namely the homogeneous contact versus the temperature, is presented in figure \ref{contact_vs_temp}. The contact is normalized to the average $k_F$ of the probed sample, and temperature is given in terms of $T/T_F$, with $T_F$ being the average Fermi temperature of the probed sample (we explain later how we determine these quantities). The data shows a monotonic decrease of the contact with increasing temperature from a maximum value of $3.3$ $Nk_F$. For $T/T_F=0.16$, at the edge of our experimentally attainable temperatures, we observe a sharp decrease of the contact to about $2.6$ $Nk_F$. In figure \ref{contact_vs_temp}, we compare our data with several theoretical models \cite{1367-2630-13-3-035007-2011} and a quantum Monte-Carlo (QMC) simulation \cite{PhysRevLett.106.205302}. The many-body theories are in the framework of the t-matrix approximation, differing by their choice of the diagrammatic expansion, the particle-particle propagator, and the self-energy. For $T/T_F>0.4$, the differences between the theoretical models are small, and the predictions all lie within the uncertainty of the data. As expected, at higher temperatures ($T/T_F>1$), we find good agreement with the virial expansions \cite{1367-2630-13-3-035007-2011} (see inset of figure \ref{contact_vs_temp}). For $T/T_F<0.4$, our data do not agree fully with any of the many-body theories. It is worth noting, however, that two of the theories (GPF and G$_0$G$_0$) predict a higher value for the contact above the superfluid phase transition than below, which may be consistent with observed sharp decrease near $T/T_F=0.16$. We note that the predicted $T_c/T_F$ has some uncertainty, as indicated by the shaded region in figure \ref{contact_vs_temp}. The non-self-consistent t-matrix model (G$_0$G$_0$) predicts an enhancement of about $50\%$ in the value of the contact around $T_c$ \cite{PhysRevA.82.021605-2010}, which the data do not show. We also do not observe an increasing trend in the contact for $T>T_c$, in contrast to a recent QMC simulation \cite{PhysRevLett.106.205302}.

As can be seen from Eq.(\ref{Tan_relation}), the contact is naturally normalized by $Nk_F$, and the detuning by the Fermi energy. However, a question which arises is how to define $E_F$ in our experiment. For a harmonically trapped gas, $E_F$ is defined in terms of the trap parameters $E_{F,\rm{trap}}=\hbar (\omega_r^2 \omega_z)^{1/3} (6 N)^{1/3}$. On the other hand, the Fermi energy of a homogeneous gas is given in terms of its density (in one spin state), $n$: $E_{F,\rm{hom}}=\frac{\hbar^2}{2m}(6\pi^2 n)^{2/3}$. In our experiment, as we increase the power of the hollow light beams, we probe a smaller portion of the gas that is more homogeneous. The relevant Fermi energy, which we use in figures \ref{raw_data} and \ref{contact_vs_temp}, is therefore the average of the local (homogeneous) Fermi energy: $E_{F,\rm{avg}}=\frac{\hbar^2}{2 m N_p}\int P(\mathbf{r}) n(\mathbf{r}) [6\pi^2 n(\mathbf{r})]^{2/3} \mathrm{d}^3 \mathrm{r}$, where $P(\mathbf{r})$ is the detection probability after optical pumping, and $N_p=\int P(\mathbf{r}) n(\mathbf{r}) \mathrm{d}^3 \mathrm{r}$ is the number of atoms probed.

The average local Fermi energy can  be obtained from the density distribution of the atoms, $n(\mathbf{r})$, and the detection probability, $P(\mathbf{r})$. We measure $n(\mathbf{r})$ by turning the trap off, without applying the hollow light beams, and imaging the cloud after 4 ms of expansion at the resonance. To determine the density distribution in trap, we fit the distribution measured after expansion and rescale the dimensions back to $t=0$, assuming hydrodynamic expansion \cite{O'Hara13122002,EPAPS_note2}. For the fit, we use the Thomas-Fermi distribution, which is known to fit the data well \cite{Ketterle2008}.

We obtain $P(\mathbf{r})$ using a model of the optical pumping by the hollow light beams \cite{PhysRevA.86.031601}. In the model, we assume that atoms that scatter a single photon are transferred to the dark state, and we account for the attenuation of the hollow light beams as they propagate through the cloud. For a given $n(\mathbf{r})$, the propagation model gives us $P(\mathbf{r})$ after the consecutive application of the two hollow light beams. We note that the results presented in figure \ref{contact_vs_temp} are not sensitive to the details of the model \cite{EPAPS_note2}.

\begin{figure}
\center\includegraphics[width=8cm]{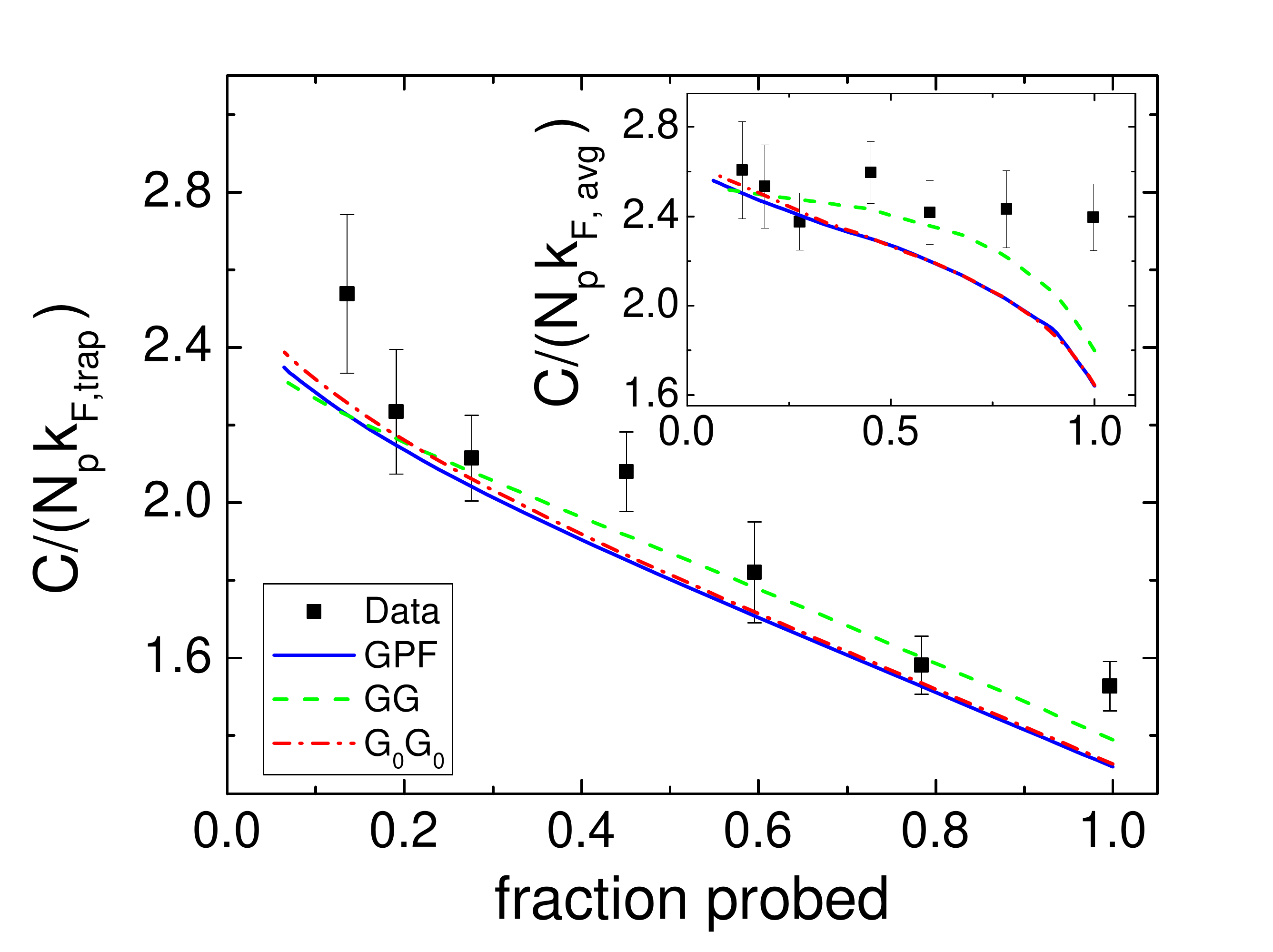}
    \caption{Contact versus the fraction of atoms probed for a gas with $T/T_F=0.46$ at the center of the cloud. In the main plot, the measured contact (squares) is normalized in respect to the trap $k_F$, and is compared to the predictions of several theoretical models (lines) using the local density approximation. The measured contact increases as we probe fewer atoms at the cloud center, where the local density is largest. The inset shows the contact normalized by the average $k_F$ of the probed atoms (squares), compared to theoretical predictions of the homogeneous contact at the average $T/T_F$ (lines).}\label{contact_vs_frac}
\end{figure}

In figure \ref{contact_vs_frac}, we show the contact at $T/T_F=0.46$ as a function of the fraction of atoms probed, which is varied by changing the intensity of the hollow light beams. The main part of figure \ref{contact_vs_frac} shows the contact per particle in units of $k_{F,\rm{trap}}$ in order to show the change in the measured signal. We find that the signal increases as we probe fewer atoms near the center of the trapped gas. We compare our results with several theoretical models, where the model lines are calculated by $C_{\rm{trap}}^{\rm{model}}=\frac{1}{N_p k_{F,\rm{trap}}}\int P(\mathbf{r}) n(\mathbf{r}) C_{\rm{hom}}^{\rm{model}}[T/T_F(\mathbf{r})] k_F(\mathbf{r}) \mathrm{d}^3 \mathrm{r}$, with $C_{\rm{hom}}^{\rm{model}}(T/T_F)$ being the model prediction for a homogeneous contact (normalized to $N k_F$), $T_F(\mathbf{r})=E_F(\mathbf{r})/k_B$ is the local Fermi temperature, and $k_B$ is the Boltzmann constant. We find good agreement of the data with the models.

In the inset of figure \ref{contact_vs_frac}, we plot the contact divided by the average local $k_F$, defined in the same way as in figure \ref{contact_vs_temp}. For comparison, we also plot theory predictions for the homogeneous contact at the average $T/T_F$,  $C_{\rm{hom}}^{\rm{model}}(\langle T/T_F\rangle)$, where the notation $\langle \rangle$ stands for density-weighted averaging. A reasonable criterion for homogeneity is when $C_{\rm{hom}}^{\rm{model}}(\langle T/T_F\rangle)\approx \langle C_{\rm{hom}}^{\rm{model}}(T/T_F)\rangle$. When the fraction of the atoms probed is less than $30\%$ we find that this approximation holds to better than $2\%$ \cite{EPAPS_note2}. When probing $30\%$ of the atoms, we calculate that the rms spread in the local $T_F$ has been reduced to about $20\%$. We find that the data for $T/T_F=0.46$ and fractions lower than $30\%$ agree with theory predictions for a homogeneous gas (see inset of figure \ref{contact_vs_frac}).

Lastly, we describe our determination of the temperature of the gas at unitarity. Thermometry of a strongly interacting gas is not trivial, and different groups have used various techniques, including thermometry with a minority component \cite{Navon07052010}, measurement of the energy versus entropy relation \cite{springerlink:10.1007/s10909-008-9850-2}, and an empirical temperature extracted from fitting the cloud to a Thomas-Fermi distribution \cite{PhysRevLett.106.170402}. We base our thermometry on a measurement of the release energy of the gas and the recently reported equation of state \cite{Ku03022012}. We determine the release energy by taking an image of cloud after $4$ ms of expansion at unitarity. Knowing our trapping potential, the equation of state, and the generalized virial theorem at unitarity \cite{springerlink:10.1007/s10909-008-9850-2}, we are left with only the temperature, $T$, as a free parameter in the calculation of the release energy. We find $T$ by matching the calculated energy to the measured one \cite{EPAPS_note2}. We estimate that the one sigma uncertainty in the temperature is $5\%$. When reporting $T/T_F$ in figure \ref{contact_vs_temp}, we use $T_F=E_{F,\rm{avg}}/k_B$.

In summary, we have presented a measurement of the homogeneous contact of a unitary Fermi gas versus temperature. Our measurement is based on a novel technique that allows us to probe local properties of the cloud. Our data show good agreement with theory predictions for $T/T_F>0.4$, but at lower temperatures no single prediction fully agrees with the data. Furthermore, the data do not show an enhanced narrow peak around T$_c$, which was predicted to exist due to pair fluctuation in a pseudogap phase. To provide additional insight into the nature of the normal state of the unitary Fermi gas, it will be interesting to test directly the pseudogap pairing instability by combining our probing technique with momentum-resolved rf spectroscopy \cite{Stewart2008,Gaebler2010}.

We acknowledge funding from the NSF and NIST.

\clearpage

\begin{widetext}


\section*{Supplementary material}

# Supplementary material for the manuscript "Measurement of the Homogeneous Contact of a Unitary Fermi gas"

Yoav Sagi, Tara E. Drake, Rabin Paudel, and Deborah S. Jin

*JILA, National Institute of Standards and Technology and the University of Colorado,*

*and the Department of Physics, University of Colorado, Boulder, CO 80309-0440, USA*

# I. SETTING THE RF POWER FOR THE LINE SHAPE MEASUREMENTS

The rf line shape is taken such that the number of atoms outcoupled by the rf pulse is small compared to total number of atoms. In order to determine the rf power that complies with this requirement, we have measured the number of atoms transferred by the rf pulse as a function of the rf power for different frequencies. The data presented here was taken without the optical pumping beams, for a gas at unitarity with $\sim 80000$ atoms per spin state and $E_f/h = 7700$ Hz, with $h$ being the Planck constant. The temperature of the gas $T/T_F = 0.13$ was measured before ramping to the resonance field.

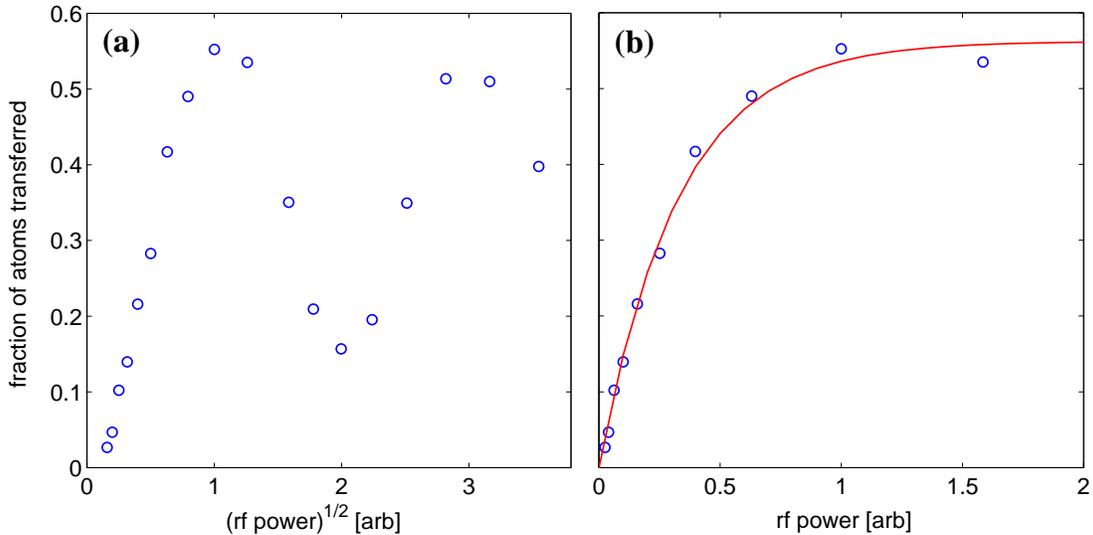

FIG. S1. **(a)** Coherent oscillations with increasing rf power at a detuning of 5 MHz relative to the single-particle transition frequency. To show the coherent oscillations, the x-axis is the square root of the rf power (which is proportional to the Rabi frequency). **(b)** Saturation with increasing rf power at a detuning of 5 MHz (zoomed in on the initial rise in figure S1, plotted as function of the rf power). The solid (red) line is the fit to the exponential model introduced in the text.

For small detunings, we observe oscillatory behavior reminiscent of coherent Rabi oscillations (figure S1a). For large detunings, the transfer is incoherent (see figure S2), and the number of outcoupled atoms saturates to a value that is between a quarter and a half of the total number of atoms (depending on the rf detuning). We fit the data to $N_{\text{out}} = A(1 - e^{-P/P_0})$ to find the saturation power $P_0$, and the results are given in Table I. For the lowest detuning, where coherent effects exist, we fit the signal up to the first peak

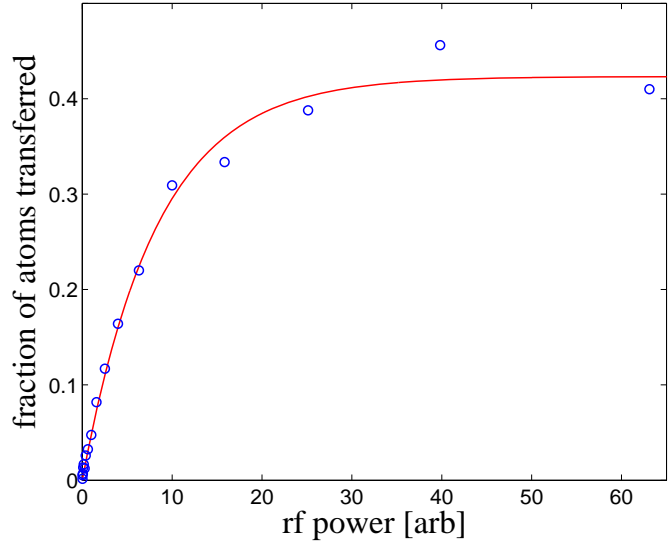

FIG. S2. Saturation with increasing rf power at a detuning of 45 MHz. The solid (red) line is the fit to the exponential model introduced in the text.

(figure S1b). When measuring the rf line shape we keep the power less than $P_0/5$, where the number of atoms transferred by the rf pulse is small compared to the total number of atoms. In this range, $N_{\text{out}} \approx A\frac{P}{P_0}$ to within 10%. We use a larger rf power at larger detunings and then linearly scale the number of atoms at each detuning to the power used for the central part of the line shape.

| detuning [kHz] | $P_0$ [arb] |
|---|---|
| 5 | 0.33 |
| 25 | 2.5 |
| 45 | 8.4 |
| 65 | 12.7 |
| 85 | 17.1 |

TABLE I. The measured saturation power, $P_0$, in arbitrary units, for different rf frequency detunings.

## II. OBTAINING THE IN-SITU DENSITY DISTRIBUTION

We use the in-situ density distribution, $n(\mathbf{r})$, in order to calculate the average $k_F$ and $E_F$ of the probed atoms. To get $n(\mathbf{r})$, we use the fact that at unitarity the cloud expands hydrodynamically, and the dynamics are governed by the continuity equation. The solution for the continuity equation with harmonic confinement with a time-dependent trapping frequency $\omega(t)$ is self-similar with the following scaling transformation: $r_i(t) = b_i(t) r_i(0)$, where $r_i$ is the spatial coordinate ($i = x, y, z$), and $b_i(t)$ obeys the equation [1, 2]:

$$\ddot{b}_i(t) = -\omega_i(t)^2 b_i(t) + \frac{\omega_i(0)^2}{b_i(t) \left[b_x(t) b_y(t) b_z(t)\right]^\gamma} \quad , \tag{1}$$

with the initial conditions $b_i(0) = 1$ and $\dot{b}_i(0) = 0$. The constant $\gamma$ is the characteristic exponent in the equation of state $\mu(n) \propto n^\gamma$, where $\mu$ is the chemical potential and $\gamma = 2/3$ at unitarity. For a sudden turn off of the trap, $\omega_i(0)$ is the trapping frequency along the $i$ axis, and $\omega_i(t > 0) = 0$.

In figure S3, we plot the measured width of the cloud in the axial and radial directions as a function of the expansion time. The data show a rapid increase in the size in the radial (tight) direction of the cloud and almost no increase in the axial direction–a characteristic of hydrodynamic expansion. The solid lines show the numerical solution of Eq.(1) with $\omega_r = 2\pi \times 226$ Hz and $\omega_z = 2\pi \times 19$ Hz (which were measured independently), which agrees very well with the data. We find that after 4 ms of expansion the finite resolution of the optical system does not affect the extracted parameters, and therefore we choose this expansion time for the density measurements. We fit the measured density profiles at 4 ms with a Thomas-Fermi distribution, which we find to be general enough for this purpose.

We have tested our density determination method by looking at the ratio of the peak density at unitarity to the peak density of a weakly interacting gas, at low temperatures. The density distribution of the weakly interacting gas is measured using a fit to a Thomas-Fermi distribution after ballistic expansion. For $T = 0$, this ratio is $n_U/n_0 = \xi^{-3/4}$, where $\xi$ at unitarity is a universal constant that relates the chemical potential to the Fermi energy: $\mu = \xi \epsilon_F$. From the measured density ratio at a temperature of $T/T_{F,\text{trap}} = 0.15$, we extract a value of $\xi = 0.40 \pm 0.05$, which is consistent with recent determinations of this universal constant by other groups [3, 4].

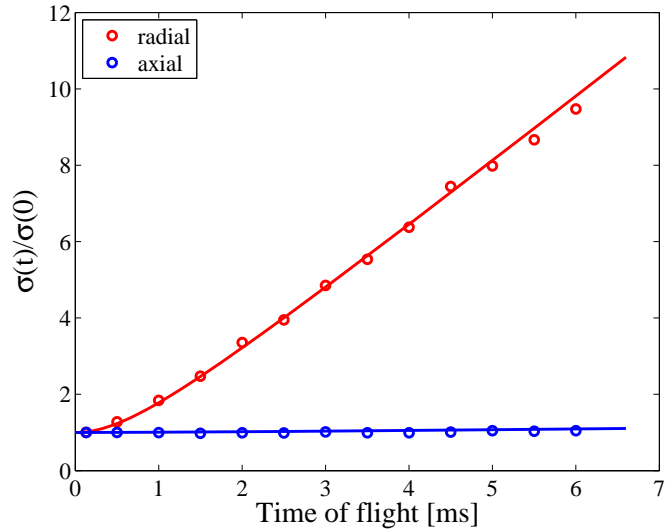

FIG. S3. Hydrodynamic expansion at unitarity. We start with a weakly interacting gas with ∼ 90000 atoms per spin state at $T/T_F = 0.12$ and ramp adiabatically to the Feshbach resonance field. We fit the cloud with a Thomas-Fermi distribution after a variable expansion time and extract the rms widths, $\sigma(t)$, in the radial and axial directions. For the data, we de-convolve the measured width with a gaussian point spread function with an rms width of 2.9 $\mu$m, to account for the finite resolution of the optical system. The data is normalized by the initial cloud size, which is 33.4 $\mu$m and 2.8 $\mu$m in the axial and radial directions, respectively. The solid lines are the numerical solution of the hydrodynamic equation.

## III. THERMOMETRY

Our thermometry assumes a knowledge of the trapping potential $V(\mathbf{r})$ and the equation of state $n(\mu, T)$, where $T$ is the temperature and $\mu$ is the chemical potential. For a non-interacting gas the equation of state is known, and for the unitary gas, we use the equation of state recently measured at MIT [3]. The trapping potential is calibrated from the known optical trap beam waists and the measured oscillation frequencies in all three directions. We adopt a local density approximation approach; the local chemical potential is given by $\mu(\mathbf{r}) = \mu_0 - V(\mathbf{r})$. For a given $T$ and number of atoms, $N$, $\mu_0$ is set by the normalization requirement $N = \int n[\mu(\mathbf{r}), T] \, \mathrm{d}^3 \mathrm{r}$. The equation of state then determines the complete density profile $n(\mathbf{r})$, from which we can calculate other quantities such as the entropy, total energy, release energy, and shape of the cloud. Since we measure $N$, we get a one-to-

one correspondence between $T$ and these quantities, and therefore any of them can serve as a thermometer. With the unitary gas, we have chosen to use the release energy as a thermometer.

We determine the release energy by measuring the cloud after it expands at unitary for 4 ms. The release energy per particle is calculated from the measured density profile of the expanded gas using

$$E_{\rm rel} = \sum_{i=x,y,z} E_{i,{\rm rel}} = \frac{1}{N} \sum_{i=x,y,z} \int \frac{m}{2}\left(\frac{r_i}{t}\right)^2 [n_t(\mathbf{r}) - n_0(\mathbf{r})] {\rm d}^3{\rm r} \ , \qquad (2)$$

where $r_i$ is the corresponding spatial coordinate ($i = x, y, z$), $t$ is the expansion time, and $n_t(\mathbf{r})$ is the density distribution at time $t$. In the experiment, we use $t = 4$ ms. We have verified that the release energy measured at $t = 4$ ms is the same as that measured after 12 ms of expansion. For a given potential $V(\mathbf{r})$, the release energy is given by [5]:

$$E_{\rm rel} = \frac{1}{2} \langle \mathbf{r} \cdot \nabla V(\mathbf{r}) \rangle \ , \qquad (3)$$

where the symbol $\langle \rangle$ stands for the density-weighted average: $\langle g(\mathbf{r}) \rangle = \frac{1}{N} \int g(\mathbf{r}) n(\mathbf{r}) {\rm d}^3{\rm r}$. By equating the calculated $E_{\rm rel}(T)$ to the measured $E_{\rm rel}$, we determine $T$.

As a comparison, we have used two other techniques to extract the temperature of the unitary gas. The first technique we compare to is based on the widely used practice of fitting the strongly interacting gas to a Thomas-Fermi distribution and extracting an empirical temperature, $\tilde{T}$, from the fitted fugacity [6]. At $T = 0$, the empirical temperature is connected to the real temperature by $T = \tilde{T}\sqrt{\xi}$, where $\xi$ is the universal constant defined above [2]. Albeit without a complete theoretical justification, one can then extend this to finite temperatures and extract $T$ [6]. In the following analysis, shown as the blue triangles in figure S4, we used $\xi = 0.376$ [3].

The second thermometry method we compare to is based on the entropy of the weakly interacting gas before the ramp to unitarity. We calculate the entropy of the weakly interacting gas from the measured temperature and the trapping potential. In the experiment, we start from the weakly interacting gas and slowly ramp to the Feshbach resonance field. By performing this ramp there and back and comparing the entropy before and after the ramp for a gas initially at $T/T_F = 0.12$ and $T/T_F = 0.22$, we have determined that the entropy increases by about 6% when going to the Feshbach resonance field. Assuming this increase, we use the entropy of the unitary gas together with the equation of state as the thermometer.

6

In figure S4, the $T/T_F$ we obtain from these two additional techniques are plotted against the release energy thermometry. We find a good agreement between all the three techniques up to $T/T_F = 0.4$. Above that temperature, the empirical temperature technique becomes unreliable since the effect of quantum degeneracy of the shape of the cloud diminishes. The entropy technique starts to show a small systematic deviation upwards above $T/T_F = 0.4$. The close agreement of the three techniques, which are based on independent observables, up to $T/T_F = 0.4$ gives us confidence in our thermometry. To estimate the errors in $T$ we look at the difference between the entropy and release energy thermometry techniques.

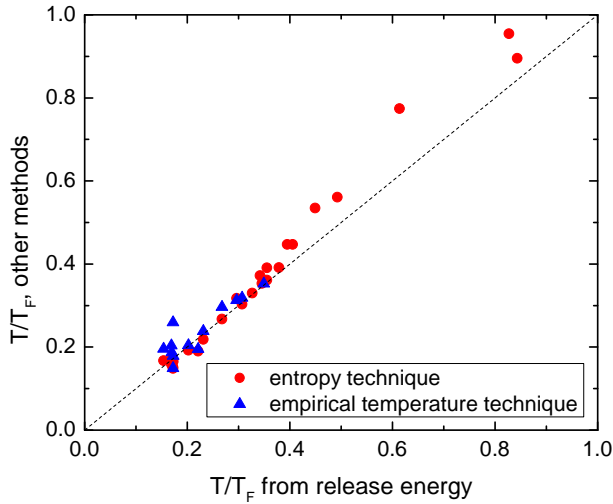

FIG. S4. Comparison of different thermometry methods. The x-axis is the temperature, $T/T_F$, where $T_F$ is the trap Fermi temperature, as extracted from the release energy. The y-axis is the temperature we get from the two other thermometry methods (see text for more details). The dashed line is $y = x$.

## IV. THE EFFECT OF THE REMAINING DENSITY INHOMOGENEITY

In figure 3 in the paper, we compare our data to the predictions of several theoretical models. Here we show that the effect of the remaining density inhomogeneity of the probed sample on the theory predictions is negligible. We use the detection probability, $P(\mathbf{r})$, and the density distribution, $n(\mathbf{r})$, for each of the data points, to calculate the average contact

7

predicted by each theoretical model according to:

$$\langle C \rangle = \frac{1}{N_p \langle k_F \rangle} \int P(\mathbf{r}) n(\mathbf{r}) C_{\text{hom}}^{\text{model}}[T/T_F(\mathbf{r})] k_F(\mathbf{r}) \mathrm{d}^3 \mathbf{r} \quad , \qquad (4)$$

where $C_{\text{hom}}^{\text{model}}$ is the prediction for the contact of a homogeneous gas theory (normalized to $Nk_F$), $N_p$ is the number of probed atoms, and $\langle k_F \rangle$ is the average $k_F$. The comparison of the average contact, $\langle C \rangle$, and the homogeneous contact for three different models is shown in figure S5. The graph clearly demonstrates that (with 30% of the atoms probed) the effect of the remaining density inhomogeneity on the contact is negligible, and hence theories for the homogeneous contact can be compared directly to the data.

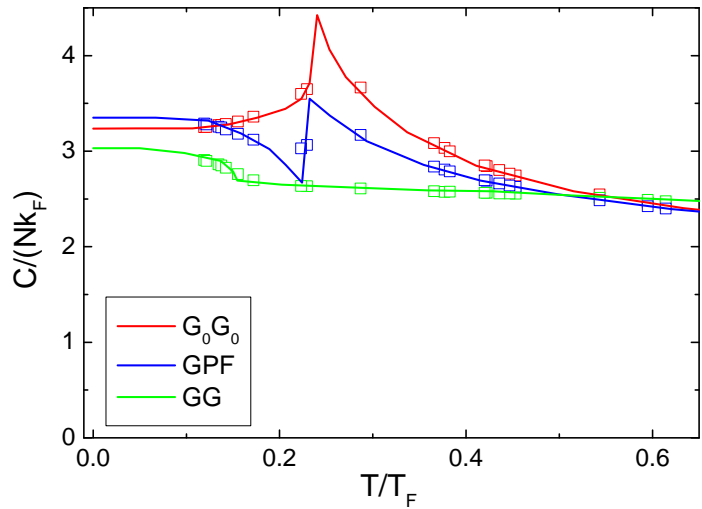

FIG. S5. Comparison of the homogeneous contact calculated by several theoretical models (solid lines) and the contact averaged over the remaining density inhomogeneity when probing the central 30% of the cloud for the same models (open symbols). The excellent agreement of the points and the lines shows that the effect of the remaining density inhomogeneity on the contact data can be neglected.

## V. THE OPTICAL PUMPING MODEL

To calculate the spatially dependent probability that atoms are probed, $P(\mathbf{r})$, we use a model for the optical pumping by the hollow light beams. In Ref. [7] we have introduced such a model, which was also tested experimentally. The calculated $P(\mathbf{r})$ along with $n(\mathbf{r})$ is

8

used in our determination of the average quantities (density, $k_F$, $T_F$, $E_F$ and the contact) of the probed atoms. We show here that the results presented in this paper are not sensitive to the details of the optical pumping model.

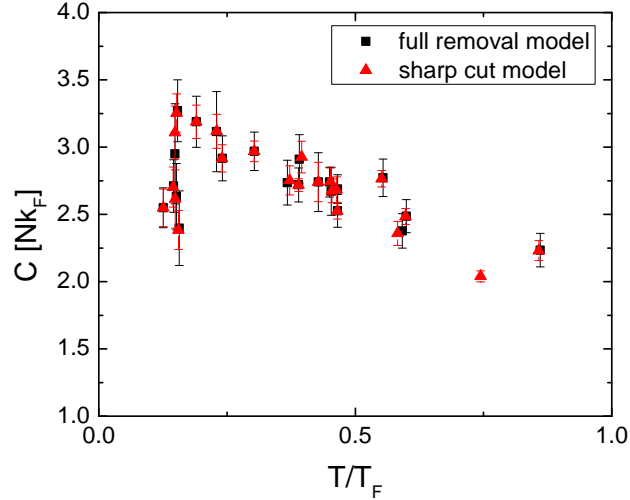

FIG. S6. The homogeneous contact at unitarity as function of the temperature, using the full removal model (squares) and the sharp cut model (triangles). The good agreement of the two models shows that the exact shape of the detection probability, $P(\mathbf{r})$, does not affect the contact data. This is because the central part of the trapped gas is nearly homogeneous, and the data was taken probing the central 30% of the atoms.

To show this, we introduce a second, much more simplified model, which we refer to as the "sharp cut" model. In the sharp cut model, we assume that atoms are left for probing only if their position $(x, y, z)$ satisfy $x^2 + y^2 < R_1^2$ and $z^2 + y^2 < R_2^2$, where $R_1$ and $R_2$ are set to reproduce the fraction of atoms probed after the application of one or both of the hollow beams. In figure S6, we present the homogeneous contact, similar to figure 3 in the paper, but using the two different models. As can be clearly seen in the figure, the results of the two models are essentially the same, which demonstrates that the details of the calculation of $P(\mathbf{r})$ are not important when the fraction of atoms probed is as small as 30%. In figure S7, we show the average Fermi energy for the data, where again the two models give consistent

9

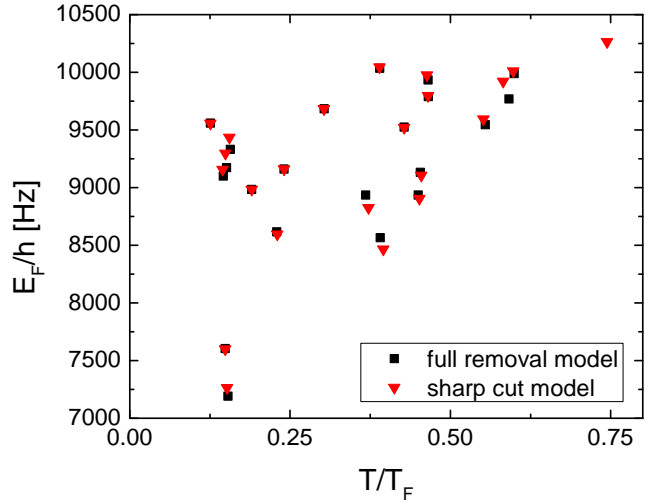

FIG. S7. The average Fermi energy with ∼ 30% of the atoms probed, using the full removal model (squares) and the sharp cut model (triangles).

results.


\end{widetext}

\end{document}